\documentclass[sigconf]{acmart}
\usepackage{geometry}
\usepackage{url}
\usepackage{multirow}
\usepackage{booktabs}
\usepackage{color}
\usepackage{hyperref}

\usepackage{tabularx} 
\usepackage{ragged2e}
\usepackage{array}
\usepackage{makecell}
\usepackage{array}
\usepackage{balance}
\newcommand{\rotatedthead}[1]{\rotatebox{90}{#1}}
\usepackage{pgfplots}
\pgfplotsset{compat=1.18}
\usetikzlibrary{patterns}
\pgfplotsset{
    width=0.95\linewidth,
    height=6cm,
    label style={font=\small},
    tick label style={font=\small},
    legend style={font=\scriptsize},
}
\acmYear{2025}
\acmConference[Submitted to RecSys 2025]{19th ACM Conference on Recommender Systems}{September 22--26,
  2025}{Prague, Czech Republic}
 
 \author{Jarne Mathi Decker}
\affiliation{%
  \institution{University of Siegen}
  \city{Siegen}
  \country{Germany}}
\email{jarne.decker@student.uni-siegen.de}

\author{Joeran Beel}
\orcid{0000-0002-4537-5573}
\affiliation{%
  \institution{University of Siegen \& Recommender-Systems.com}
  \city{Siegen}
  \country{Germany}}  
\email{joeran.beel@uni-siegen.de}

\title[RecSys Meta-Learning via Algorithm Features]{
Algorithm Selection for Recommender Systems via Meta-Learning on Algorithm Characteristics
}
\begin{document}
\begin{abstract}
The Algorithm Selection Problem for recommender systems—choosing the best algorithm for a given user or context—remains a significant challenge. Traditional meta-learning approaches often treat algorithms as categorical choices, ignoring their intrinsic properties. Recent work has shown that explicitly characterizing algorithms with features can improve model performance in other domains. Building on this, we propose a per-user meta-learning approach for recommender system selection that leverages both user meta-features and automatically extracted algorithm features from source code. Our preliminary results, averaged over six diverse datasets, show that augmenting a meta-learner with algorithm features improves its average NDCG@10 performance by 8.83\% from 0.135 (user features only) to 0.147. This enhanced model outperforms the Single Best Algorithm baseline (0.131) and successfully closes 10.5\% of the performance gap to a theoretical oracle selector. These findings show that even static source code metrics provide a valuable predictive signal, presenting a promising direction for building more robust and intelligent recommender systems.
\end{abstract}



\maketitle

\section{Introduction}
It is a well-established principle in recommender systems and machine learning that no single algorithm performs optimally across all problem types, a manifestation of the "No Free Lunch" theorem \cite{wolpert}. In the domain of recommender systems, this is particularly evident: evaluations regularly show that the performance of recommendation algorithms are inconsistent across different applications or datasets \cite{beelreproducibility}. This performance variance extends down to the individual user level, where the optimal algorithm can differ significantly between users within the same context \cite{collins,ekstrand}. 

\subsection{Background: Algorithm Selection}
The high context dependence of algorithm performance leads to the Algorithm Selection Problem (ASP) \cite{Rice}. The standard practice of selecting the Single Best Algorithm (\textbf{SBA}) for a diverse user base is inherently suboptimal. It leads to a quantifiable performance loss for a large fraction of users for whom the chosen algorithm is not the best fit. At least in theory, it would be ideal to identify and then use the individually best algorithm for each user, item, or scenario. To do this perfectly well, an "Oracle" would be needed \cite{tornedeoracle}. The performance gap between the Single Best Algorithm (SBA) and such a theoretical Oracle selector - which could choose the best algorithm for each user - is substantial \cite{perinstance2020}. Consequently, a common approach across various fields is to employ a meta-learning paradigm to approximate this Oracle, thereby automating algorithm selection for contexts like recommender systems \cite{cunha, wegmeth, collins, ekstrand, kerschke, kotthoff}. Here, for each instance in a dataset a machine learning model (meta-learner) is trained to predict which algorithm will perform best on that individual instance. 

\subsection{Research Problem}
In the past years, there has been lots of research on algorithm selection for recommender systems \cite{cunha,  ekstrand}, also, but not exclusively, from our own research group \cite{Anand2020,Arabzadeh2024,Arambakam2020,Beel2019,Beel2017,Beel2015,Beel2019a,Beel2019b,Beel2020,Beel2019c,Beel2018,Beel2017a,Collins2018,Collins2018a,Edenhofer2019,Gupta2020,Vente2023,Vente2024,Vente2024a,Vente2023a,Vente2022,Wegmeth2023,Wegmeth2022,Wegmeth2022a,Wegmeth2023a,wegmeth,collins, beel}.
However, current works on meta-learning, especially in the domain of recommender systems, treat algorithms as featureless classes. This means, based on e.g. user, item or dataset characteristics, the meta-learner tries to learn which algorithm is best, without knowing anything about the algorithm itself.

Recognizing the limitations of treating algorithms as feature-less black boxes, recent research has begun to explore the use of algorithm features ($f_A$) to explicitly characterize the algorithms themselves. The primary motivation is to improve the prediction quality and to train a single, unified meta-model that can generalize to new, unseen algorithms by understanding their properties~\cite{cenikj, pulatov}. To the best of our knowledge, there are few researchers who have explored this (Table \ref{tab:algo_feature_overview}). These feature types range from static, implementation-level properties like source code metrics~\cite{pulatov} to more abstract, semantic representations derived from Large Language Models~\cite{wu} or Knowledge Graphs (KG) ~\cite{kostovska23}. Other approaches characterize algorithms by their configurable parameters~\cite{tornede} or their empirical performance footprints on a set of probe datasets~\cite{eftimov}.

While this prior work
establishes the value of $f_A$, its application context is critical. Research on algorithm features has
largely focused on domains such as SAT solving \cite{pulatov, wu},
continuous optimization \cite{eftimov, kostovska23, kostovska} or dataset-level selection of
ML classifiers \cite{tornede}.

\subsection{Research Goal}

The application of this feature-based meta-learning approach to the per-user algorithm selection task in recommender systems remains an open and promising field of research. Standard meta-learning approaches for recommender systems still largely treat diverse algorithms like k-NN and Matrix Factorization as equivalent "choices," despite their fundamentally different operational principles~\cite{ekstrand}. To the best of our knowledge, a systematic study combining automatically extracted algorithm features with user features for this specific task has not been conducted. Our work aims to fill this gap by providing a first empirical analysis using static source code metrics as a practical and fully automated source of algorithm features.

Our key hypothesis is that by making the meta-learner aware of the fundamental differences between algorithms—via explicit algorithm features ($f_A$)—its selection capabilities can be improved beyond what is possible using only user characteristics ($f_I$). For example, a meta-learner could learn that users with sparse interaction histories ($f_I$) benefit from algorithms with low code complexity ($f_A$), a connection that is impossible to learn without an explicit algorithm representation.
Our goal, hence, is to explore if and to what extent the use of algorithm features can improve meta-learning for recommender systems algorithm selection.

\begin{table}[h]
\caption{Overview of Algorithm Feature ($f_A$) Generation Approaches from Literature.}
\vspace{-5pt}
\label{tab:algo_feature_overview}
\small
\begin{tabularx}{\linewidth}{@{}l >{\RaggedRight}X >{\RaggedRight\arraybackslash}X@{}}
\toprule
\textbf{Feature Type} & \textbf{Key Paper(s)} & \textbf{Brief Description}\\
\midrule
Source Code \& AST & Pulatov et al. \cite{pulatov} & Static analysis (LOC, CC) and AST graph properties. \\
\addlinespace
LLM Embeddings & Wu et al. \cite{wu} & High-dimensional embeddings from code/text via LLMs. \\
\addlinespace
Explainability & Kostovska et al. \cite{kostovska23}, Nikolikj et al. \cite{nikolikj} & Vector of SHAP scores of instance features. \\
\addlinespace
KG Embeddings & Kostovska et al. \cite{kostovska23} & Learned vectors of algorithm nodes from a KG. \\
\addlinespace
Hyperparameters & Tornede et al. \cite{tornede} & Vector of hyperparameter values defining a configuration. \\
\addlinespace
Performance & Eftimov et al. \cite{eftimov} & Vector of performance scores on benchmark problems. \\
\bottomrule
\end{tabularx}
\end{table}
\vspace{-10pt}

\section{Methodology}
%/ So eine Übersicht brauchen wir nicht bei einem so kurzen Kapitel.  Die folgenden Unterkapitel sollten so selbsterklärend sein, dass der Zusammenhang unmittelbar klar wird.

%/Auskommentiert: To test our hypothesis, we designed a multi-stage experimental pipeline. The process involves (1) generating a robust ground-truth dataset of per-user algorithm performances, (2) engineering distinct feature sets to characterize both users and algorithms, and (3) training and evaluating our meta-learning models against established baselines using a rigorous cross-validation scheme.

\subsection{Ground Truth Generation}
\textbf{Datasets}: We conducted all experiments on Six datasets from different domains, and with different data characteristics: MovieLens-1M, LastFM-360K, Book-Crossing, RetailRocket, Steam and a dataset containing restaurant-ratings. After applying initial preprocessing, such as filtering users with fewer than 10 interactions, the final statistics of these datasets are summarized in Table \ref{tab:dataset_stats}.

\textbf{Algorithms portfolio}: Our portfolio comprises nine implementations from LensKit and RecBole, selected to cover a range of distinct algorithmic paradigms. We include classic baselines (Popularity), neighborhood-based methods (ItemKNN), matrix factorization (BPR, ImplicitMF), an autoencoder-based model (EASE), and a sequential model (FPMC). To assess potential implementation-specific effects, our portfolio includes versions of Poularity, ItemKNN, and BPR from both libraries.

\textbf{HPO}: For the sake of time and energy efficiency, we did not optimize hyperparameters for the single algorithms. For our scenario, we consider this acceptable as our goal is not to develop one single best algorithm but to predict which algorithm out of a pool of algorithms performs best. This is regardless of hyperparameters.   

\textbf{Performance and Ground Truth}; We applied a temporal evaluation protocol. For each user in every dataset, we sorted their interaction history chronologically and split it into an 80\%/20\% training/test split. Each algorithm from the portfolio was  trained on the aggregate training data of all users. We measured performance by nDCG@10.

This process yielded our ground truth: a performance matrix $P_{u,a}$ where each entry represents the NDCG@10 score for user $u$ with algorithm $a$.

\begin{table}[h]
\caption{Statistics of Preprocessed Datasets.}
\vspace{-5pt}
\label{tab:dataset_stats}
\small
\begin{tabular}{lrrrr}
\toprule
\textbf{Dataset} & \textbf{Users} & \textbf{Items} & \textbf{Interactions} & \textbf{Sparsity} \\
\midrule
MovieLens    & 6,040  & 3,706  & 1,000,209  & 95.53\% \\
LastFM       & 1,874  & 17,612 & 92,779   & 99.72\% \\
BookCrossing & 2,946 & 17,384 & 272,677  & 99.47\% \\
RetailRocket & 9,446  & 68,433 & 240,843  & 99.96\% \\
Steam        & 2,189  & 5,076  & 104,737   & 99.06\% \\
Restaurants & 59 & 84 & 681 & 86.26\% \\
\bottomrule
\end{tabular}
\end{table}

\subsection{Meta-Feature Engineering}
For meta-learning, we engineered features describing the users ($f_I$) and algorithms ($f_A$).  

\textbf{User Features ($f_I$):} We represent each user as a vector of 15 meta-features derived from their training data. These features capture multiple dimensions including activity (e.g. number of interactions), rating patterns (e.g. average rating and rating entropy), temporal dynamics (e.g. history duration), and item popularity preferences (e.g. average popularity of all items the user interacted with). For our experiments, these features cover the standard approach for Algorithm Selection, where only instances, in this case users, are explicitly characterized through features.

\textbf{Algorithm Features ($f_A$):} 
As an initial, automatically extractable representation, we characterized each algorithm implementation using static source code analysis via the Radon tool \footnote{Radon: A Python tool for computing code metrics. Available at: \url{https://radon.readthedocs.io/}}. This provided a quantitative "fingerprint" of each implementation based on metrics for size (e.g. SLOC), complexity (e.g. Average Cyclomatic Complexity), and Halstead metrics (e.g., Effort).

In addition, we constructed the Abstract Syntax Tree for each algorithm implementation and calculated AST-features like node count, average degree and depth using the networkx library.
A full list of the features $f_A$ we utilized is provided in Table \ref{tab:algo_meta_feature_list_compact}.

\begin{table}[htbp]
\centering
\caption{Overview of Algorithm Meta-Features ($f_A$).}
\vspace{-5pt}
\label{tab:algo_meta_feature_list_compact}
\begin{tabular}{@{}ll@{}}
\toprule
\textbf{Feature Name} & \textbf{Category} \\
\midrule
\multicolumn{2}{l}{\textit{Static \& Complexity Metrics}} \\
\quad \texttt{sloc} & Size \\
\quad \texttt{lloc} & Size \\
\quad \texttt{average\_cc\_file} & Complexity \\
\quad \texttt{num\_complexity\_blocks} & Complexity \\
\addlinespace
\multicolumn{2}{l}{\textit{Halstead Metrics}} \\
\quad \texttt{hal\_volume} & Code Volume \\
\quad \texttt{hal\_difficulty} & Code Difficulty \\
\quad \texttt{hal\_effort} & Implementation Effort \\
\addlinespace
\multicolumn{2}{l}{\textit{AST Graph Metrics}} \\
\quad \texttt{ast\_node\_count} & Graph Structure \\
\quad \texttt{ast\_edge\_count} & Graph Structure \\
\quad \texttt{ast\_avg\_degree} & Graph Structure \\
\quad \texttt{ast\_max\_degree} & Graph Structure \\
\quad \texttt{ast\_transitivity} & Graph Structure \\
\quad \texttt{ast\_avg\_clustering} & Graph Structure \\
\quad \texttt{ast\_depth} & Graph Structure \\
\bottomrule
\end{tabular}
\end{table}

\subsection{Meta-Learning and Evaluation}
We designed two meta-learning models and evaluated them against standard baselines to quantify their effectiveness.

\textbf{Baselines:} We established two performance baselines. The Single Best Algorithm (SBA) represents the performance of the single best-performing algorithm when applied uniformly to all users in a dataset. The Virtual Best Algorithm (VBA), or Oracle, represents the theoretical maximum performance achievable by a perfect per-user algorithm selector that chooses the best algorithm for each user.

\textbf{Meta-Learners:} We compare two models:
\begin{enumerate}
    \item \textit{M(User-Only)}: A baseline meta-learner trained only on user features ($f_I$) to predict a performance vector for all candidate algorithms. This would represent the state of the art in meta-learning for recommender systems.
    \item \textit{M(User+Algo)}: Our proposed model trained on a restructured dataset of (user, algorithm) pairs. It uses a concatenated features vector of both users and algorithm features ($f_I,f_A$) to predict a single performance score.
\end{enumerate}
For this initial work, we used a hyperparameter-tuned LightGBM regressor as the underlying model for both approaches to ensure a fair comparison.

\textbf{Evaluation Protocol} To obtain robust performance estimates, we employ a 5-fold cross-validation scheme. The data is split on \texttt{user\_id} to ensure that all interactions from a single user remain in the same fold, preventing data leakage. For each fold, we train our meta-learners on the training users and evaluate them on the held-out test users. The final reported scores are the average across all 5 folds. We report three metrics:
(1) \textbf{Avg. NDCG@10}, the average actual quality of the selected algorithms; (2) \textbf{Top-1 Accuracy}, the percentage of times the single best algorithm is correctly identified; and (3) \textbf{Top-3 Accuracy}, the percentage of times the actual best algorithm is among the top three predicted choices.

\section{Results}
\label{results}
\subsection{Baselines}
When choosing the single best algorithms for each dataset, an average nDCG@10 of 0.131 is achieved (Table \ref{tab:final_results_summary}). The Oracle - the theoretical algorithm selector that chooses the best algorithm for each user - could theoretically achieve an NDCG of 0.282 (+116\%). The theoretical performance of the Oracle demonstrates the potential of algorithm selection.

%/we present the main performance comparison. The first step in our analysis is to quantify the potential for improvement. A significant performance gap exists between the Single Best Algorithm (SBA) and the Virtual Best Algorithm(VBA) across all datasets. On average, the VBA achieves an NDCG@10 of 0.2820, which is a 116\% relative improvement over the SBA performance of 0.1306. This substantial gap confirms that a per-user algorithm selection strategy holds significant potential to enhance overall system performance.

\subsection{Meta-Learner Performance}
Our baseline meta-learner \textit{M (User-Only)} achieved an NDCG of 0.135 on average over all datasets. This is a slight improvement over using the single best algorithms. Also, the meta learner achieved an accuracy of 20.24\% in predicting the best performing algorithm. Accuracy for predicting one of the top 3 algorithms was 59.07\%, respectively. This indicates that user features alone contain a predictive signal, though precision in identifying the single best algorithm is limited.

\begin{figure}[htbp]
\centering
\begin{tikzpicture}
\begin{axis}[
    ybar,
    bar width=20pt,
    enlarge x limits=0.9,
    ylabel={Average NDCG@10},
    symbolic x coords={SBA, M (User-Only), M (User+Algo), VBA (Oracle)},
    xtick=data,
    ymin=0,
    enlarge y limits={value=0.09, upper},
    xticklabels={}, 
    tick style={draw=none}, 
    nodes near coords,
    nodes near coords align={vertical},
    nodes near coords style={font=\small},
    legend pos=north west,
    legend cell align={left}
]

\addplot coordinates {(SBA, 0.131)};
\addplot coordinates {(M (User-Only), 0.135)};
\addplot coordinates {(M (User+Algo), 0.147)};
\addplot coordinates {(VBA (Oracle), 0.282)};

\legend{SBA, M (User-Only), M (User+Algo), VBA (Oracle)}
\end{axis}
\end{tikzpicture}
\vspace{-15pt}
\caption{Average performance (NDCG@10) across all six datasets.}
\label{fig:perf_summary_average}
\end{figure}
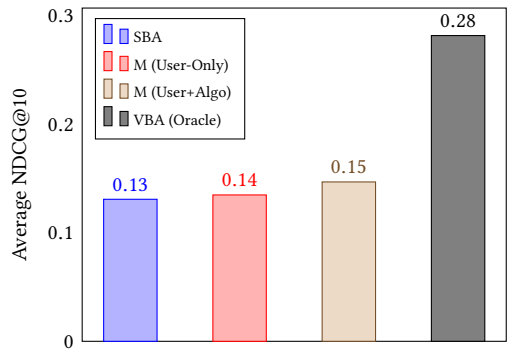

Our novel \textit{M (User+Algo)} model, augmented with algorithm features, achieved an average nDCG of 0.147. This model outperforms the SBA on four of the six datasets, with an average performance gain of 12.07\%.
Its average Top-1 Accuracy was 21.63\%, and its Top-3 Accuracy was 62.74\%.
In Figure \ref{fig:perf_summary_average} we provide a visual summary of these results, comparing the average performance of both meta-learners against the SBA and VBS baselines across all datasets.
\begin{table*}[htbp]
\centering
\small 
\caption{Performance Summary: Baselines vs. Meta-Learners across all Datasets (Metric: Avg. NDCG@10)}
\label{tab:final_results_summary}
\setlength{\tabcolsep}{3pt}

\begin{tabular}{@{}l l r r | r r r | r r r r r r@{}}
\toprule
& & \multicolumn{2}{c|}{\textbf{Baselines}} & \multicolumn{3}{c|}{\textbf{M (User Features Only)}} & \multicolumn{6}{c}{\textbf{M (User + Algorithm Features)}} \\
\cmidrule(lr){3-4} \cmidrule(lr){5-7} \cmidrule(lr){8-13}
\textbf{Dataset} & \textbf{SB Algorithm} & 
\rotatedthead{SBA Perf.} & 
\rotatedthead{VBA Perf.} & 
\rotatedthead{Perf.} & 
\rotatedthead{Acc. @1} & 
\rotatedthead{Acc. @3} & 
\rotatedthead{Perf.} & 
\rotatedthead{\% Gain / SBA} & 
\rotatedthead{\% Gain / User ML} & 
\rotatedthead{Acc. @1} & 
\rotatedthead{Acc. @3} & 
\rotatedthead{\% Gap Closed} \\
\midrule
MovieLens    & LK\_BPR       & 0.284 & 0.616 & 0.331 & 23.39\% & 48.43\% & 0.332 & +16.99\% & +0.27\% & 19.69\% & 48.79\% & 14.54\% \\
LastFM       & LK\_BPR       & 0.038 & 0.086 & 0.049 & 28.28\% & 88.42\% & 0.052 & +38.56\% & +5.89\% & 23.33\% & 93.92\% & 29.90\% \\
BookCrossing & RB\_ItemKNN   & 0.041 & 0.072 & 0.037 & 11.98\% & 77.12\% & 0.040 & -3.89\% & +6.18\% & 11.03\% & 92.26\% & -4.97\% \\
RetailRocket & LK\_ImplicitMF & 0.107 & 0.181 & 0.107 & 13.84\% & 43.18\% & 0.106 & -0.84\% & -1.30\% & 12.72\% & 40.06\% & -1.26\% \\
Steam        & LK\_BPR       & 0.163 & 0.358 & 0.200 & 34.76\% & 65.14\% & 0.195 & +19.16\% & -2.50\% & 42.53\% & 70.81\% & 16.10\% \\
Restaurants  & LK\_ImplicitMF & 0.150 & 0.380 & 0.083 & 13.33\% & 32.12\% & 0.154 & +2.46\% & +86.56\% & 20.45\% & 30.61\% & 1.64\% \\
\midrule
\textbf{Average} & \textbf{N/A} & \textbf{0.131} & \textbf{0.282} & \textbf{0.135} & \textbf{20.24\%} & \textbf{59.07\%} & \textbf{0.147} & \textbf{+12.07\%} & \textbf{+8.83\%} & \textbf{21.63\%} & \textbf{62.74\%} & \textbf{10.49\%} \\
\bottomrule
\end{tabular}
\\
\scriptsize{\textbf{Note:} SBA = Single Best Algorithm; VBA = Virtual Best Algorithm (Oracle); M = Meta-Learner. `Perf.` columns show average NDCG@10. `Acc. @1` / `Acc. @3` refer to Top-1 and Top-3 selection accuracy. `\% Gap Closed` = `(ML Perf. - SBA Perf.) / (VBA Perf. - SBA Perf.)`.}
\end{table*}

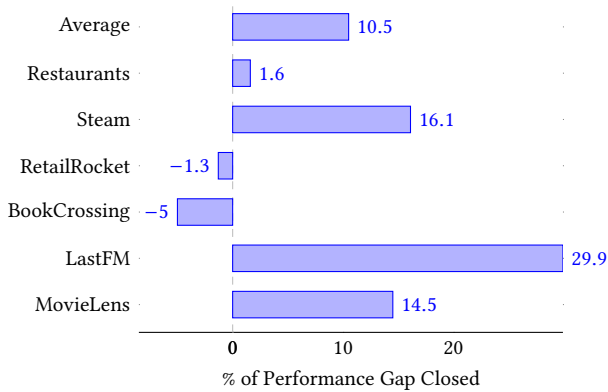
\begin{figure}[h!]
\centering
\begin{tikzpicture}
\begin{axis}[
    xbar,
    y axis line style = { opacity = 0 },
    axis x line* = bottom,
    tickwidth         = 0pt,
    enlarge x limits={value=0.1, upper}, 
    enlarge x limits={value=0.1, lower},   
    xlabel={\% of Performance Gap Closed},
    symbolic y coords={MovieLens, LastFM, BookCrossing, RetailRocket, Steam, Restaurants, Average},
    ytick             = data,
    nodes near coords,
    nodes near coords align={horizontal},
    nodes near coords style={
        font=\small,
        /pgf/number format/fixed, 
        /pgf/number format/precision=1
    },
    extra x ticks={0},
    extra x tick style={grid=major, black, dashed},
]
\addplot coordinates {(14.5,MovieLens) (29.9,LastFM) (-5.0,BookCrossing) (-1.3,RetailRocket) (16.1,Steam) (1.6,Restaurants) (10.5,Average)};
\end{axis}
\end{tikzpicture}
\vspace{-15pt}
\caption{Percentage of the performance gap between the SBA and the VBA (Oracle) that was closed by the M (User+Algo) meta-learner.}
\label{fig:gap_closed}
\end{figure}

\subsection{The Impact of Algorithm Features}
Our primary hypothesis is that explicitly modeling algorithm characteristics improves selection performance. The results in Table \ref{tab:final_results_summary} allow for a direct comparison between the two meta-learners.

Comparing the two meta-learners reveals that the effect of algorithm features is highly dependent on dataset characteristics. We observed a positive performance gain on four of the six datasets. This effect was most pronounced on the Restaurants dataset (+86.56\%), where the baseline \textit{M (User-Only)} model performed particularly poorly, suggesting that the algorithm features provided a crucial signal that was absent in the user features for that specific data structure. On larger datasets, the performance gains were more moderate and even slightly negative on the Steam and RetailRocket datasets.

Relative to the baselines, the \textit{M (User+Algo)} model closed an average of 10.49\% of the total performance gap between the SBA and the VBA (\% Gap Closed). The effectiveness varies significantly by dataset, with the strongest results on LastFM (29.90\% gap closed) and Steam (16.10\%). In Figure \ref{fig:gap_closed} we visualize this variance, highlighting the datasets where the meta-learner was most and least effective at closing the performance gap relative to the SBS. The negative gap closure on some datasets indicates that for certain data characteristics, the current set of source code and AST-features is not yet sufficient to consistently outperform the strong SBA, highlighting a clear direction for future  feature engineering.

\section{Conclusion and Future Work}
In this work, we presented preliminary results from a meta-learning framework for per-user recommender system selection. Our empirical evaluation across six diverse datasets demonstrates that a meta-learner, when augmented with a combination of user and algorithm features, can consistently outperform a strong Single Best Algorithm baseline. We showed that adding algorithm features provides a tangible, though moderate, improvement in both average recommendation quality (NDCG@10) and selection accuracy over a meta-learner that relies on user features alone.

These findings provide a strong foundation for future research. Our immediate next steps are to:
\begin{itemize}
    \item Expand the Algorithm Feature Set: To address cases where our current model underperformed, we will incorporate more diverse features, particularly perfor-mance-based landmarkers \cite{eftimov} and potentially manually engineered conceptual features (e.g., algorithm family, learning paradigm) to provide a richer signal to the meta-learner. 
    \item Diversify Models and Datasets: We plan to expand our algorithm portfolio with more diverse paradigms and continue to add datasets to further test the generalizability of our approach.
    \item Explore Advanced Meta-Learners: We will investigate more advanced model architectures, such as Factorization Machines or Two-Tower Neural Networks, which are explicitly designed to model the interactions between user and algorithm feature sets.
\end{itemize}
This work successfully demonstrates the promise of a feature-based approach for algorithm selection in recommender systems. The next phase of our research will build directly on this foundation, shifting from static code analysis to behavioral landmarking features and from standard regressors to interaction-aware architectures, in pursuit of a meta-learning model truly aware of algorithm characteristics and behavior.

\bibliographystyle{ACM-Reference-Format}
\bibliography{references}


\begin{thebibliography}{45}

%%% the Web address.

\ifx \showCODEN    \undefined \def \showCODEN     #1{\unskip}     \fi
\ifx \showDOI      \undefined \def \showDOI       #1{#1}\fi
\ifx \showISBNx    \undefined \def \showISBNx     #1{\unskip}     \fi
\ifx \showISBNxiii \undefined \def \showISBNxiii  #1{\unskip}     \fi
\ifx \showISSN     \undefined \def \showISSN      #1{\unskip}     \fi
\ifx \showLCCN     \undefined \def \showLCCN      #1{\unskip}     \fi
\ifx \shownote     \undefined \def \shownote      #1{#1}          \fi
\ifx \showarticletitle \undefined \def \showarticletitle #1{#1}   \fi
\ifx \showURL      \undefined \def \showURL       {\relax}        \fi
% invisible to TeX
\providecommand\bibfield[2]{#2}
\providecommand\bibinfo[2]{#2}
\providecommand\natexlab[1]{#1}
\providecommand\showeprint[2][]{arXiv:#2}

\bibitem[Anand and Beel(2020)]%
        {Anand2020}
\bibfield{author}{\bibinfo{person}{Rohan Anand} {and} \bibinfo{person}{Joeran Beel}.} \bibinfo{year}{2020}\natexlab{}.
\newblock \showarticletitle{Auto-Surprise: An Automated Recommender-System (AutoRecSys) Library with Tree of Parzens Estimator (TPE) Optimization}. In \bibinfo{booktitle}{\emph{14th ACM Conference on Recommender Systems (RecSys)}}. \bibinfo{pages}{1--4}.
\newblock
\urldef\tempurl%
\url{https://arxiv.org/abs/2008.13532}
\showURL{%
\tempurl}


\bibitem[Arabzadeh et~al\mbox{.}(2024)]%
        {Arabzadeh2024}
\bibfield{author}{\bibinfo{person}{Ardalan Arabzadeh}, \bibinfo{person}{Tobias Vente}, {and} \bibinfo{person}{Joeran Beel}.} \bibinfo{year}{2024}\natexlab{}.
\newblock \showarticletitle{Green Recommender Systems: Optimizing Dataset Size for Energy-Efficient Algorithm Performance}. In \bibinfo{booktitle}{\emph{International Workshop on Recommender Systems for Sustainability and Social Good (RecSoGood) at the 18th ACM Conference on Recommender Systems (ACM RecSys)}}.
\newblock
\urldef\tempurl%
\url{https://isg.beel.org/pubs/2024-Green-RecSys-Dataset-Sampling-Ardalan.pdf}
\showURL{%
\tempurl}


\bibitem[Arambakam and Beel(2020)]%
        {Arambakam2020}
\bibfield{author}{\bibinfo{person}{Mukesh Arambakam} {and} \bibinfo{person}{Joeran Beel}.} \bibinfo{year}{2020}\natexlab{}.
\newblock \showarticletitle{Federated Meta-Learning: Democratizing Algorithm Selection Across Disciplines and Software Libraries}. In \bibinfo{booktitle}{\emph{7th ICML Workshop on Automated Machine Learning}}. \bibinfo{pages}{1--8}.
\newblock
\urldef\tempurl%
\url{https://www.automl.org/wp-content/uploads/2020/07/AutoML_2020_paper_39.pdf}
\showURL{%
\tempurl}


\bibitem[Beel(2015)]%
        {Beel2015}
\bibfield{author}{\bibinfo{person}{Joeran Beel}.} \bibinfo{year}{2015}\natexlab{}.
\newblock \showarticletitle{Towards Effective Research-Paper Recommender Systems and User Modeling based on Mind Maps}.
\newblock \bibinfo{journal}{\emph{PhD Thesis. Otto-von-Guericke Universit{\"a}t Magdeburg}} (\bibinfo{year}{2015}).
\newblock


\bibitem[Beel(2017)]%
        {Beel2017}
\bibfield{author}{\bibinfo{person}{Joeran Beel}.} \bibinfo{year}{2017}\natexlab{}.
\newblock \showarticletitle{A Macro/Micro Recommender System for Recommendation Algorithms [Proposal]}.
\newblock \bibinfo{journal}{\emph{ResearchGate}} (\bibinfo{year}{2017}).
\newblock
\urldef\tempurl%
\url{https://doi.org/10.13140/RG.2.2.14978.79047}
\showDOI{\tempurl}


\bibitem[Beel(2019)]%
        {Beel2019}
\bibfield{author}{\bibinfo{person}{Joeran Beel}.} \bibinfo{year}{2019}\natexlab{}.
\newblock \showarticletitle{Federated Meta-Learning: Democratizing Algorithm Selection Across Disciplines and Software Libraries (Proposal)}.
\newblock \bibinfo{journal}{\emph{ResearchGate}} (\bibinfo{year}{2019}).
\newblock
\urldef\tempurl%
\url{https://doi.org/10.13140/RG.2.2.25744.35844}
\showDOI{\tempurl}


\bibitem[Beel et~al\mbox{.}(2017)]%
        {Beel2017a}
\bibfield{author}{\bibinfo{person}{Joeran Beel}, \bibinfo{person}{Akiko Aizawa}, \bibinfo{person}{Corinna Breitinger}, {and} \bibinfo{person}{Bela Gipp}.} \bibinfo{year}{2017}\natexlab{}.
\newblock \showarticletitle{Mr. DLib: Recommendations-as-a-service (RaaS) for Academia}. In \bibinfo{booktitle}{\emph{Proceedings of the 17th ACM/IEEE Joint Conference on Digital Libraries}} (Toronto, Ontario, Canada) \emph{(\bibinfo{series}{JCDL '17})}. \bibinfo{publisher}{IEEE Press}, \bibinfo{address}{Piscataway, NJ, USA}, \bibinfo{pages}{313--314}.
\newblock
\showISBNx{978-1-5386-3861-3}
\urldef\tempurl%
\url{http://dl.acm.org/citation.cfm?id=3200334.3200389}
\showURL{%
\tempurl}


\bibitem[Beel et~al\mbox{.}(2016)]%
        {beelreproducibility}
\bibfield{author}{\bibinfo{person}{Joeran Beel}, \bibinfo{person}{Corinna Breitinger}, \bibinfo{person}{Stefan Langer}, \bibinfo{person}{Andreas Lommatzsch}, {and} \bibinfo{person}{Bela Gipp}.} \bibinfo{year}{2016}\natexlab{}.
\newblock \showarticletitle{Towards reproducibility in recommender-systems research}.
\newblock \bibinfo{journal}{\emph{User modeling and user-adapted interaction}}  \bibinfo{volume}{26} (\bibinfo{year}{2016}), \bibinfo{pages}{69--101}.
\newblock


\bibitem[Beel et~al\mbox{.}(2018)]%
        {Beel2018}
\bibfield{author}{\bibinfo{person}{Joeran Beel}, \bibinfo{person}{Andrew Collins}, {and} \bibinfo{person}{Akiko Aizawa}.} \bibinfo{year}{2018}\natexlab{}.
\newblock \showarticletitle{The Architecture of Mr. DLib’s Scientific Recommender-System API}. In \bibinfo{booktitle}{\emph{Proceedings of the 26th Irish Conference on Artificial Intelligence and Cognitive Science (AICS)}}, Vol.~\bibinfo{volume}{2259}. \bibinfo{publisher}{CEUR-WS}, \bibinfo{pages}{78--89}.
\newblock


\bibitem[Beel et~al\mbox{.}(2019)]%
        {Beel2019c}
\bibfield{author}{\bibinfo{person}{Joeran Beel}, \bibinfo{person}{Alan Griffin}, {and} \bibinfo{person}{Conor O'Shey}.} \bibinfo{year}{2019}\natexlab{}.
\newblock \showarticletitle{Darwin \& Goliath: Recommendations-As-a-Service with Automated Algorithm-Selection and White-Labels}. In \bibinfo{booktitle}{\emph{13th ACM Conference on Recommender Systems (RecSys)}}.
\newblock


\bibitem[Beel and Kotthoff(2019a)]%
        {Beel2019b}
\bibfield{author}{\bibinfo{person}{Joeran Beel} {and} \bibinfo{person}{Lars Kotthoff}.} \bibinfo{year}{2019}\natexlab{a}.
\newblock \showarticletitle{Preface: The 1st Interdisciplinary Workshop on Algorithm Selection and Meta-Learning in Information Retrieval (AMIR)}. In \bibinfo{booktitle}{\emph{Proceddings of The 1st Interdisciplinary Workshop on Algorithm Selection and Meta-Learning in Information Retrieval (AMIR)}}. \bibinfo{pages}{1--9}.
\newblock


\bibitem[Beel and Kotthoff(2019b)]%
        {Beel2019a}
\bibfield{author}{\bibinfo{person}{Joeran Beel} {and} \bibinfo{person}{Lars Kotthoff}.} \bibinfo{year}{2019}\natexlab{b}.
\newblock \showarticletitle{Proposal for the 1st Interdisciplinary Workshop on Algorithm Selection and Meta-Learning in Information Retrieval (AMIR)}. In \bibinfo{booktitle}{\emph{Proceedings of the 41st European Conference on Information Retrieval (ECIR)}} \emph{(\bibinfo{series}{Lecture Notes in Computer Science book series (LNCS)}, Vol.~\bibinfo{volume}{11438})}, \bibfield{editor}{\bibinfo{person}{L.~Azzopardi}, \bibinfo{person}{B.~Stein}, \bibinfo{person}{N.~Fuhr}, \bibinfo{person}{P.~Mayr}, \bibinfo{person}{C.~Hauff}, {and} \bibinfo{person}{D.~Hiemstra}} (Eds.). \bibinfo{publisher}{Springer}, \bibinfo{pages}{383--388}.
\newblock
\urldef\tempurl%
\url{https://doi.org/10.1007/978-3-030-15719-7_53}
\showDOI{\tempurl}


\bibitem[Beel et~al\mbox{.}(2020)]%
        {Beel2020}
\bibfield{author}{\bibinfo{person}{Joeran Beel}, \bibinfo{person}{Bryan Tyrell}, \bibinfo{person}{Edward Bergman}, \bibinfo{person}{Andrew Collins}, {and} \bibinfo{person}{Shahad Nagoor}.} \bibinfo{year}{2020}\natexlab{}.
\newblock \showarticletitle{Siamese Meta-Learning and Algorithm Selection with ‘Algorithm-Performance Personas’ [Proposal]}.
\newblock \bibinfo{journal}{\emph{arXiv:2006.12328 [cs.LG]}} (\bibinfo{year}{2020}).
\newblock


\bibitem[Cenikj et~al\mbox{.}(2024)]%
        {cenikj}
\bibfield{author}{\bibinfo{person}{Gjorgjina Cenikj}, \bibinfo{person}{Ana Nikolikj}, \bibinfo{person}{Gašper Petelin}, \bibinfo{person}{Niki van Stein}, \bibinfo{person}{Carola Doerr}, {and} \bibinfo{person}{Tome Eftimov}.} \bibinfo{year}{2024}\natexlab{}.
\newblock \bibinfo{title}{A Survey of Meta-features Used for Automated Selection of Algorithms for Black-box Single-objective Continuous Optimization}.
\newblock
\newblock
\showeprint[arxiv]{2406.06629}~[cs.LG]
\urldef\tempurl%
\url{https://arxiv.org/abs/2406.06629}
\showURL{%
\tempurl}


\bibitem[Collins and Beel(2019)]%
        {collins}
\bibfield{author}{\bibinfo{person}{Andrew Collins} {and} \bibinfo{person}{Joeran Beel}.} \bibinfo{year}{2019}\natexlab{}.
\newblock \bibinfo{title}{Meta-Learned Per-Instance Algorithm Selection in Scholarly Recommender Systems}.
\newblock
\newblock
\showeprint[arxiv]{1912.08694}~[cs.IR]
\urldef\tempurl%
\url{https://arxiv.org/abs/1912.08694}
\showURL{%
\tempurl}


\bibitem[Collins et~al\mbox{.}(2020)]%
        {perinstance2020}
\bibfield{author}{\bibinfo{person}{Andrew Collins}, \bibinfo{person}{Laura Tierney}, {and} \bibinfo{person}{Joeran Beel}.} \bibinfo{year}{2020}\natexlab{}.
\newblock \bibinfo{title}{Per-Instance Algorithm Selection for Recommender Systems via Instance Clustering}.
\newblock
\newblock
\urldef\tempurl%
\url{https://doi.org/10.48550/arXiv.2012.15151}
\showDOI{\tempurl}


\bibitem[Collins et~al\mbox{.}(2018a)]%
        {Collins2018}
\bibfield{author}{\bibinfo{person}{Andrew Collins}, \bibinfo{person}{Dominika Tkaczyk}, {and} \bibinfo{person}{Joeran Beel}.} \bibinfo{year}{2018}\natexlab{a}.
\newblock \showarticletitle{A Novel Approach to Recommendation Algorithm Selection using Meta-Learning}. In \bibinfo{booktitle}{\emph{Proceedings of the 26th Irish Conference on Artificial Intelligence and Cognitive Science (AICS)}}, Vol.~\bibinfo{volume}{2259}. \bibinfo{publisher}{CEUR-WS}, \bibinfo{pages}{210--219}.
\newblock


\bibitem[Collins et~al\mbox{.}(2018b)]%
        {Collins2018a}
\bibfield{author}{\bibinfo{person}{Andrew Collins}, \bibinfo{person}{Dominika Tkaczyk}, {and} \bibinfo{person}{Joeran Beel}.} \bibinfo{year}{2018}\natexlab{b}.
\newblock \showarticletitle{One-at-a-time: A Meta-Learning Recommender-System for Recommendation-Algorithm Selection on Micro Level}. In \bibinfo{booktitle}{\emph{arXiv:1805.12118}}.
\newblock


\bibitem[Cunha et~al\mbox{.}(2018)]%
        {cunha}
\bibfield{author}{\bibinfo{person}{Tiago Cunha}, \bibinfo{person}{Carlos Soares}, {and} \bibinfo{person}{Andr{\'e}~CPLF de Carvalho}.} \bibinfo{year}{2018}\natexlab{}.
\newblock \showarticletitle{Metalearning and Recommender Systems: A literature review and empirical study on the algorithm selection problem for Collaborative Filtering}.
\newblock \bibinfo{journal}{\emph{Information Sciences}}  \bibinfo{volume}{423} (\bibinfo{year}{2018}), \bibinfo{pages}{128--144}.
\newblock


\bibitem[Edenhofer et~al\mbox{.}(2019)]%
        {Edenhofer2019}
\bibfield{author}{\bibinfo{person}{Gordian Edenhofer}, \bibinfo{person}{Andrew Collins}, \bibinfo{person}{Akiko Aizawa}, {and} \bibinfo{person}{Joeran Beel}.} \bibinfo{year}{2019}\natexlab{}.
\newblock \showarticletitle{Augmenting the DonorsChoose.org Corpus for Meta-Learning}. In \bibinfo{booktitle}{\emph{Proceedings of The 1st Interdisciplinary Workshop on Algorithm Selection and Meta-Learning in Information Retrieval (AMIR)}}. \bibinfo{publisher}{CEUR-WS}, \bibinfo{pages}{32--38}.
\newblock


\bibitem[Eftimov et~al\mbox{.}(2020)]%
        {eftimov}
\bibfield{author}{\bibinfo{person}{Tome Eftimov}, \bibinfo{person}{Gorjan Popovski}, \bibinfo{person}{Dragi Kocev}, {and} \bibinfo{person}{Peter Koro\v{s}ec}.} \bibinfo{year}{2020}\natexlab{}.
\newblock \showarticletitle{Performance2vec: a step further in explainable stochastic optimization algorithm performance}. In \bibinfo{booktitle}{\emph{Proceedings of the 2020 Genetic and Evolutionary Computation Conference Companion}} (Canc\'{u}n, Mexico) \emph{(\bibinfo{series}{GECCO '20})}. \bibinfo{publisher}{Association for Computing Machinery}, \bibinfo{address}{New York, NY, USA}, \bibinfo{pages}{193–194}.
\newblock
\showISBNx{9781450371278}
\urldef\tempurl%
\url{https://doi.org/10.1145/3377929.3390020}
\showDOI{\tempurl}


\bibitem[Ekstrand and Riedl(2012)]%
        {ekstrand}
\bibfield{author}{\bibinfo{person}{Michael Ekstrand} {and} \bibinfo{person}{John Riedl}.} \bibinfo{year}{2012}\natexlab{}.
\newblock \showarticletitle{When recommenders fail: predicting recommender failure for algorithm selection and combination}. In \bibinfo{booktitle}{\emph{Proceedings of the Sixth ACM Conference on Recommender Systems}} (Dublin, Ireland) \emph{(\bibinfo{series}{RecSys '12})}. \bibinfo{publisher}{Association for Computing Machinery}, \bibinfo{address}{New York, NY, USA}, \bibinfo{pages}{233–236}.
\newblock
\showISBNx{9781450312707}
\urldef\tempurl%
\url{https://doi.org/10.1145/2365952.2366002}
\showDOI{\tempurl}


\bibitem[Gupta and Beel(2020)]%
        {Gupta2020}
\bibfield{author}{\bibinfo{person}{Srijan Gupta} {and} \bibinfo{person}{Joeran Beel}.} \bibinfo{year}{2020}\natexlab{}.
\newblock \showarticletitle{Auto-CaseRec: Automatically Selecting and Optimizing Recommendation-Systems Algorithms}.
\newblock \bibinfo{journal}{\emph{OSF Preprints DOI:10.31219/osf.io/4znmd,}} (\bibinfo{year}{2020}).
\newblock
\urldef\tempurl%
\url{https://doi.org/10.31219/osf.io/4znmd}
\showDOI{\tempurl}


\bibitem[Kerschke et~al\mbox{.}(2018)]%
        {kerschke}
\bibfield{author}{\bibinfo{person}{Pascal Kerschke}, \bibinfo{person}{Holger~H. Hoos}, \bibinfo{person}{Frank Neumann}, {and} \bibinfo{person}{Heike Trautmann}.} \bibinfo{year}{2018}\natexlab{}.
\newblock \bibinfo{title}{Automated Algorithm Selection: Survey and Perspectives}.
\newblock
\newblock
\showeprint[arxiv]{1811.11597}~[cs.LG]
\urldef\tempurl%
\url{https://arxiv.org/abs/1811.11597}
\showURL{%
\tempurl}


\bibitem[Kostovska et~al\mbox{.}(2022)]%
        {kostovska}
\bibfield{author}{\bibinfo{person}{Ana Kostovska}, \bibinfo{person}{Diederick Vermetten}, \bibinfo{person}{Sa\v{s}o D\v{z}eroski}, \bibinfo{person}{Carola Doerr}, \bibinfo{person}{Peter Korosec}, {and} \bibinfo{person}{Tome Eftimov}.} \bibinfo{year}{2022}\natexlab{}.
\newblock \showarticletitle{The importance of landscape features for performance prediction of modular CMA-ES variants}. In \bibinfo{booktitle}{\emph{Proceedings of the Genetic and Evolutionary Computation Conference}} (Boston, Massachusetts) \emph{(\bibinfo{series}{GECCO '22})}. \bibinfo{publisher}{Association for Computing Machinery}, \bibinfo{address}{New York, NY, USA}, \bibinfo{pages}{648–656}.
\newblock
\showISBNx{9781450392372}
\urldef\tempurl%
\url{https://doi.org/10.1145/3512290.3528832}
\showDOI{\tempurl}


\bibitem[Kostovska et~al\mbox{.}(2023)]%
        {kostovska23}
\bibfield{author}{\bibinfo{person}{Ana Kostovska}, \bibinfo{person}{Diederick Vermetten}, \bibinfo{person}{Sa{\v{s}}o D{\v{z}}eroski}, \bibinfo{person}{Pan{\v{c}}e Panov}, \bibinfo{person}{Tome Eftimov}, {and} \bibinfo{person}{Carola Doerr}.} \bibinfo{year}{2023}\natexlab{}.
\newblock \showarticletitle{Using knowledge graphs for performance prediction of modular optimization algorithms}. In \bibinfo{booktitle}{\emph{International Conference on the Applications of Evolutionary Computation (Part of EvoStar)}}. Springer, \bibinfo{pages}{253--268}.
\newblock


\bibitem[Kotthoff(2016)]%
        {kotthoff}
\bibfield{author}{\bibinfo{person}{Lars Kotthoff}.} \bibinfo{year}{2016}\natexlab{}.
\newblock \showarticletitle{Algorithm selection for combinatorial search problems: A survey}.
\newblock In \bibinfo{booktitle}{\emph{Data mining and constraint programming: Foundations of a cross-disciplinary approach}}. \bibinfo{publisher}{Springer}, \bibinfo{pages}{149--190}.
\newblock


\bibitem[Nikolikj et~al\mbox{.}(2022)]%
        {nikolikj}
\bibfield{author}{\bibinfo{person}{Ana Nikolikj}, \bibinfo{person}{Ryan Lang}, \bibinfo{person}{Peter Koro{\v{s}}ec}, {and} \bibinfo{person}{Tome Eftimov}.} \bibinfo{year}{2022}\natexlab{}.
\newblock \showarticletitle{Explaining Differential Evolution Performance Through Problem Landscape Characteristics}. In \bibinfo{booktitle}{\emph{Bioinspired Optimization Methods and Their Applications}}, \bibfield{editor}{\bibinfo{person}{Marjan Mernik}, \bibinfo{person}{Tome Eftimov}, {and} \bibinfo{person}{Matej {\v{C}}repin{\v{s}}ek}} (Eds.). \bibinfo{publisher}{Springer International Publishing}, \bibinfo{address}{Cham}, \bibinfo{pages}{99--113}.
\newblock
\showISBNx{978-3-031-21094-5}


\bibitem[Pulatov et~al\mbox{.}(2022)]%
        {pulatov}
\bibfield{author}{\bibinfo{person}{Damir Pulatov}, \bibinfo{person}{Marie Anastacio}, \bibinfo{person}{Lars Kotthoff}, {and} \bibinfo{person}{Holger Hoos}.} \bibinfo{year}{2022}\natexlab{}.
\newblock \showarticletitle{Opening the Black Box: Automated Software Analysis for Algorithm Selection}. In \bibinfo{booktitle}{\emph{Proceedings of the First International Conference on Automated Machine Learning}} \emph{(\bibinfo{series}{Proceedings of Machine Learning Research}, Vol.~\bibinfo{volume}{188})}, \bibfield{editor}{\bibinfo{person}{Isabelle Guyon}, \bibinfo{person}{Marius Lindauer}, \bibinfo{person}{Mihaela van~der Schaar}, \bibinfo{person}{Frank Hutter}, {and} \bibinfo{person}{Roman Garnett}} (Eds.). \bibinfo{publisher}{PMLR}, \bibinfo{pages}{6/1--18}.
\newblock
\urldef\tempurl%
\url{https://proceedings.mlr.press/v188/pulatov22a.html}
\showURL{%
\tempurl}


\bibitem[Rice(1976)]%
        {Rice}
\bibfield{author}{\bibinfo{person}{John~R. Rice}.} \bibinfo{year}{1976}\natexlab{}.
\newblock \showarticletitle{The Algorithm Selection Problem**This work was partially supported by the National Science Foundation through Grant GP-32940X. This chapter was presented as the George E. Forsythe Memorial Lecture at the Computer Science Conference, February 19, 1975, Washington, D. C.}
\newblock \bibinfo{series}{Advances in Computers}, Vol.~\bibinfo{volume}{15}. \bibinfo{publisher}{Elsevier}, \bibinfo{pages}{65--118}.
\newblock
\showISSN{0065-2458}
\urldef\tempurl%
\url{https://doi.org/10.1016/S0065-2458(08)60520-3}
\showDOI{\tempurl}


\bibitem[Tornede et~al\mbox{.}(2023)]%
        {tornedeoracle}
\bibfield{author}{\bibinfo{person}{Alexander Tornede}, \bibinfo{person}{Lukas Gehring}, \bibinfo{person}{Tanja Tornede}, \bibinfo{person}{Marcel Wever}, {and} \bibinfo{person}{Eyke H{\"u}llermeier}.} \bibinfo{year}{2023}\natexlab{}.
\newblock \showarticletitle{Algorithm selection on a meta level}.
\newblock \bibinfo{journal}{\emph{Machine Learning}} \bibinfo{volume}{112}, \bibinfo{number}{4} (\bibinfo{year}{2023}), \bibinfo{pages}{1253--1286}.
\newblock


\bibitem[Tornede et~al\mbox{.}(2020)]%
        {tornede}
\bibfield{author}{\bibinfo{person}{Alexander Tornede}, \bibinfo{person}{Marcel Wever}, {and} \bibinfo{person}{Eyke Hüllermeier}.} \bibinfo{year}{2020}\natexlab{}.
\newblock \bibinfo{booktitle}{\emph{Extreme Algorithm Selection with Dyadic Feature Representation}}.
\newblock \bibinfo{publisher}{Springer International Publishing}, \bibinfo{pages}{309–324}.
\newblock
\showISBNx{9783030615277}
\showISSN{1611-3349}
\urldef\tempurl%
\url{https://doi.org/10.1007/978-3-030-61527-7_21}
\showDOI{\tempurl}


\bibitem[Tyrrell et~al\mbox{.}(2020)]%
        {beel}
\bibfield{author}{\bibinfo{person}{Bryan Tyrrell}, \bibinfo{person}{Edward Bergman}, \bibinfo{person}{Gareth Jones}, {and} \bibinfo{person}{Joeran Beel}.} \bibinfo{year}{2020}\natexlab{}.
\newblock \showarticletitle{Algorithm-Performance Personas’ for Siamese Meta-Learning and Automated Algorithm Selection}. In \bibinfo{booktitle}{\emph{7th ICML Workshop on Automated Machine Learning}}, Vol.~\bibinfo{volume}{1}. \bibinfo{pages}{16}.
\newblock


\bibitem[Vente(2023)]%
        {Vente2023}
\bibfield{author}{\bibinfo{person}{Tobias Vente}.} \bibinfo{year}{2023}\natexlab{}.
\newblock \showarticletitle{Advancing Automation of Design Decisions in Recommender System Pipelines}. In \bibinfo{booktitle}{\emph{Proceedings of the 17th ACM Conference on Recommender Systems}}. \bibinfo{pages}{1355--1360}.
\newblock
\urldef\tempurl%
\url{https://doi.org/doi/10.1145/3604915.3608886}
\showDOI{\tempurl}


\bibitem[Vente and Beel(2024)]%
        {Vente2024}
\bibfield{author}{\bibinfo{person}{Tobias Vente} {and} \bibinfo{person}{Joeran Beel}.} \bibinfo{year}{2024}\natexlab{}.
\newblock \showarticletitle{The Potential of AutoML for Recommender Systems}.
\newblock \bibinfo{journal}{\emph{arXiv}} (\bibinfo{year}{2024}), \bibinfo{pages}{18}.
\newblock
\urldef\tempurl%
\url{https://doi.org/10.48550/arXiv.2402.04453}
\showDOI{\tempurl}


\bibitem[Vente et~al\mbox{.}(2023)]%
        {Vente2023a}
\bibfield{author}{\bibinfo{person}{Tobias Vente}, \bibinfo{person}{Michael Ekstrand}, {and} \bibinfo{person}{Joeran Beel}.} \bibinfo{year}{2023}\natexlab{}.
\newblock \showarticletitle{Introducing LensKit-Auto, an Experimental Automated Recommender System (AutoRecSys) Toolkit}. In \bibinfo{booktitle}{\emph{Proceedings of the 17th ACM Conference on Recommender Systems}}. \bibinfo{pages}{1212--1216}.
\newblock
\urldef\tempurl%
\url{https://dl.acm.org/doi/10.1145/3604915.3610656}
\showURL{%
\tempurl}


\bibitem[Vente et~al\mbox{.}(2024)]%
        {Vente2024a}
\bibfield{author}{\bibinfo{person}{Tobias Vente}, \bibinfo{person}{Zainil Mehta}, \bibinfo{person}{Lukas Wegmeth}, {and} \bibinfo{person}{Joeran Beel}.} \bibinfo{year}{2024}\natexlab{}.
\newblock \showarticletitle{Greedy Ensemble Selection for Top-N Recommendations}. In \bibinfo{booktitle}{\emph{RobustRecSys Workshop at the 18th ACM Conference on Recommender Systems (ACM RecSys)}}.
\newblock


\bibitem[Vente et~al\mbox{.}(2022)]%
        {Vente2022}
\bibfield{author}{\bibinfo{person}{Tobias Vente}, \bibinfo{person}{Lennart Purucker}, {and} \bibinfo{person}{Joeran Beel}.} \bibinfo{year}{2022}\natexlab{}.
\newblock \showarticletitle{The Feasibility of Greedy Ensemble Selection for Automated Recommender Systems}.
\newblock \bibinfo{journal}{\emph{COSEAL Workshop 2022}} (\bibinfo{year}{2022}).
\newblock
\urldef\tempurl%
\url{http://dx.doi.org/10.13140/RG.2.2.16277.29921}
\showURL{%
\tempurl}


\bibitem[Wegmeth(2023)]%
        {Wegmeth2023}
\bibfield{author}{\bibinfo{person}{Lukas Wegmeth}.} \bibinfo{year}{2023}\natexlab{}.
\newblock \showarticletitle{Improving Recommender Systems Through the Automation of Design Decisions}. In \bibinfo{booktitle}{\emph{Proceedings of the 17th ACM Conference on Recommender Systems}}. \bibinfo{pages}{1332--1338}.
\newblock
\urldef\tempurl%
\url{https://dl.acm.org/doi/pdf/10.1145/3604915.3608877}
\showURL{%
\tempurl}


\bibitem[Wegmeth and Beel(2022a)]%
        {Wegmeth2022}
\bibfield{author}{\bibinfo{person}{Lukas Wegmeth} {and} \bibinfo{person}{Joeran Beel}.} \bibinfo{year}{2022}\natexlab{a}.
\newblock \showarticletitle{CaMeLS: Cooperative Meta-Learning Service for Recommender Systems}. In \bibinfo{booktitle}{\emph{Proceedings of the 2nd Perspectives on the Evaluation of Recommender Systems Workshop}}.
\newblock
\urldef\tempurl%
\url{https://ceur-ws.org/Vol-3228/paper2.pdf}
\showURL{%
\tempurl}


\bibitem[Wegmeth and Beel(2022b)]%
        {Wegmeth2022a}
\bibfield{author}{\bibinfo{person}{Lukas Wegmeth} {and} \bibinfo{person}{Joeran Beel}.} \bibinfo{year}{2022}\natexlab{b}.
\newblock \showarticletitle{Cooperative Meta-Learning Service for Recommender Systems}.
\newblock \bibinfo{journal}{\emph{COSEAL Workshop 2022}} (\bibinfo{year}{2022}).
\newblock
\urldef\tempurl%
\url{http://dx.doi.org/10.13140/RG.2.2.10667.41768}
\showURL{%
\tempurl}


\bibitem[Wegmeth et~al\mbox{.}(2023)]%
        {Wegmeth2023a}
\bibfield{author}{\bibinfo{person}{Lukas Wegmeth}, \bibinfo{person}{Tobias Vente}, {and} \bibinfo{person}{Joeran Beel}.} \bibinfo{year}{2023}\natexlab{}.
\newblock \showarticletitle{The Challenges of Algorithm Selection and Hyperparameter Optimization for Recommender Systems}.
\newblock \bibinfo{journal}{\emph{COSEAL Workshop 2023}} (\bibinfo{year}{2023}).
\newblock
\urldef\tempurl%
\url{http://dx.doi.org/10.13140/RG.2.2.24089.19049}
\showURL{%
\tempurl}


\bibitem[Wegmeth et~al\mbox{.}(2024)]%
        {wegmeth}
\bibfield{author}{\bibinfo{person}{Lukas Wegmeth}, \bibinfo{person}{Tobias Vente}, {and} \bibinfo{person}{Joeran Beel}.} \bibinfo{year}{2024}\natexlab{}.
\newblock \showarticletitle{Recommender Systems Algorithm Selection for Ranking Prediction on Implicit Feedback Datasets}. In \bibinfo{booktitle}{\emph{Proceedings of the 18th ACM Conference on Recommender Systems}}. \bibinfo{pages}{1163--1167}.
\newblock


\bibitem[Wolpert and Macready(1997)]%
        {wolpert}
\bibfield{author}{\bibinfo{person}{D.H. Wolpert} {and} \bibinfo{person}{W.G. Macready}.} \bibinfo{year}{1997}\natexlab{}.
\newblock \showarticletitle{No free lunch theorems for optimization}.
\newblock \bibinfo{journal}{\emph{IEEE Transactions on Evolutionary Computation}} \bibinfo{volume}{1}, \bibinfo{number}{1} (\bibinfo{year}{1997}), \bibinfo{pages}{67--82}.
\newblock
\urldef\tempurl%
\url{https://doi.org/10.1109/4235.585893}
\showDOI{\tempurl}


\bibitem[Wu et~al\mbox{.}(2023)]%
        {wu}
\bibfield{author}{\bibinfo{person}{Xingyu Wu}, \bibinfo{person}{Yan Zhong}, \bibinfo{person}{Jibin Wu}, \bibinfo{person}{Bingbing Jiang}, {and} \bibinfo{person}{Kay~Chen Tan}.} \bibinfo{year}{2023}\natexlab{}.
\newblock \showarticletitle{Large language model-enhanced algorithm selection: towards comprehensive algorithm representation}.
\newblock \bibinfo{journal}{\emph{arXiv preprint arXiv:2311.13184}} (\bibinfo{year}{2023}).
\newblock


\end{thebibliography}
\end{document}